\documentclass[twoside,twocolumn,9pt]{article}
\usepackage{extsizes}
\usepackage[super,sort&compress,comma]{natbib} 
\usepackage[version=3]{mhchem}
\usepackage[left=1.5cm, right=1.5cm, top=1.785cm, bottom=2.0cm]{geometry}
\usepackage{balance}
\usepackage{mathptmx}
\usepackage{sectsty}
\usepackage{graphicx} 
\usepackage{lastpage}
\usepackage[format=plain,justification=justified,singlelinecheck=false,font={stretch=1.125,small,sf},labelfont=bf,labelsep=space]{caption}
\usepackage{float}
\usepackage{fancyhdr}
\usepackage{fnpos}
\usepackage[english]{babel}
\addto{\captionsenglish}{%
  
}
\usepackage{array}
\usepackage{droidsans}
\usepackage{charter}
\usepackage[T1]{fontenc}
\usepackage[usenames,dvipsnames]{xcolor}
\usepackage{setspace}
\usepackage[compact]{titlesec}
\usepackage{hyperref}
\usepackage{epstopdf}%This line makes .eps figures into .pdf - please comment out if not required.
\usepackage{amssymb}
\definecolor{cream}{RGB}{222,217,201}

\begin{document}
\graphicspath{{Figures/}}
\pagestyle{fancy}
\thispagestyle{plain}
\fancypagestyle{plain}{
%%%HEADER%%%
\renewcommand{\headrulewidth}{0pt}
}
%%%END OF HEADER%%%

%%%PAGE SETUP - Please do not change any commands within this section%%%
\makeFNbottom
\makeatletter
\renewcommand\LARGE{\@setfontsize\LARGE{15pt}{17}}
\renewcommand\Large{\@setfontsize\Large{12pt}{14}}
\renewcommand\large{\@setfontsize\large{10pt}{12}}
\renewcommand\footnotesize{\@setfontsize\footnotesize{7pt}{10}}
\makeatother

\renewcommand{\thefootnote}{\fnsymbol{footnote}}
\renewcommand\footnoterule{\vspace*{1pt}% 
\color{cream}\hrule width 3.5in height 0.4pt \color{black}\vspace*{5pt}} 
\setcounter{secnumdepth}{5}

\makeatletter 
\renewcommand\@biblabel[1]{#1}            
\renewcommand\@makefntext[1]% 
{\noindent\makebox[0pt][r]{\@thefnmark\,}#1}
\makeatother 
\renewcommand{\figurename}{\small{Fig.}~}
\sectionfont{\sffamily\Large}
\subsectionfont{\normalsize}
\subsubsectionfont{\bf}
\setstretch{1.125} %In particular, please do not alter this line.
\setlength{\skip\footins}{0.8cm}
\setlength{\footnotesep}{0.25cm}
\setlength{\jot}{10pt}
\titlespacing*{\section}{0pt}{4pt}{4pt}
\titlespacing*{\subsection}{0pt}{15pt}{1pt}
%%%END OF PAGE SETUP%%%

%%%FOOTER%%%
\fancyfoot{}
\fancyfoot[LO,RE]{\vspace{-7.1pt}\includegraphics[height=9pt]{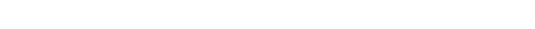}}
\fancyfoot[CO]{\vspace{-7.1pt}\hspace{13.2cm}\includegraphics{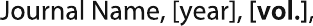}}
\fancyfoot[CE]{\vspace{-7.2pt}\hspace{-14.2cm}\includegraphics{head_foot/RF}}
\fancyfoot[RO]{\footnotesize{\sffamily{1--\pageref{LastPage} ~\textbar  \hspace{2pt}\thepage}}}
\fancyfoot[LE]{\footnotesize{\sffamily{\thepage~\textbar\hspace{3.45cm} 1--\pageref{LastPage}}}}
\fancyhead{}
\renewcommand{\headrulewidth}{0pt} 
\renewcommand{\footrulewidth}{0pt}
\setlength{\arrayrulewidth}{1pt}
\setlength{\columnsep}{6.5mm}
\setlength\bibsep{1pt}
%%%END OF FOOTER%%%

%%%FIGURE SETUP - please do not change any commands within this section%%%
\makeatletter 
\newlength{\figrulesep} 
\setlength{\figrulesep}{0.5\textfloatsep} 

\newcommand{\topfigrule}{\vspace*{-1pt}% 
\noindent{\color{cream}\rule[-\figrulesep]{\columnwidth}{1.5pt}} }

\newcommand{\botfigrule}{\vspace*{-2pt}% 
\noindent{\color{cream}\rule[\figrulesep]{\columnwidth}{1.5pt}} }

\newcommand{\dblfigrule}{\vspace*{-1pt}% 
\noindent{\color{cream}\rule[-\figrulesep]{\textwidth}{1.5pt}} }

\makeatother
%%%END OF FIGURE SETUP%%%

%%%TITLE, AUTHORS AND ABSTRACT%%%
\twocolumn[
  \begin{@twocolumnfalse}
{\includegraphics[height=30pt]{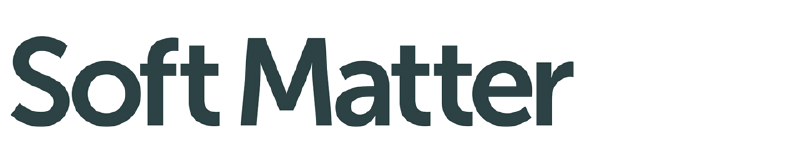}\hfill\raisebox{0pt}[0pt][0pt]{\includegraphics[height=55pt]{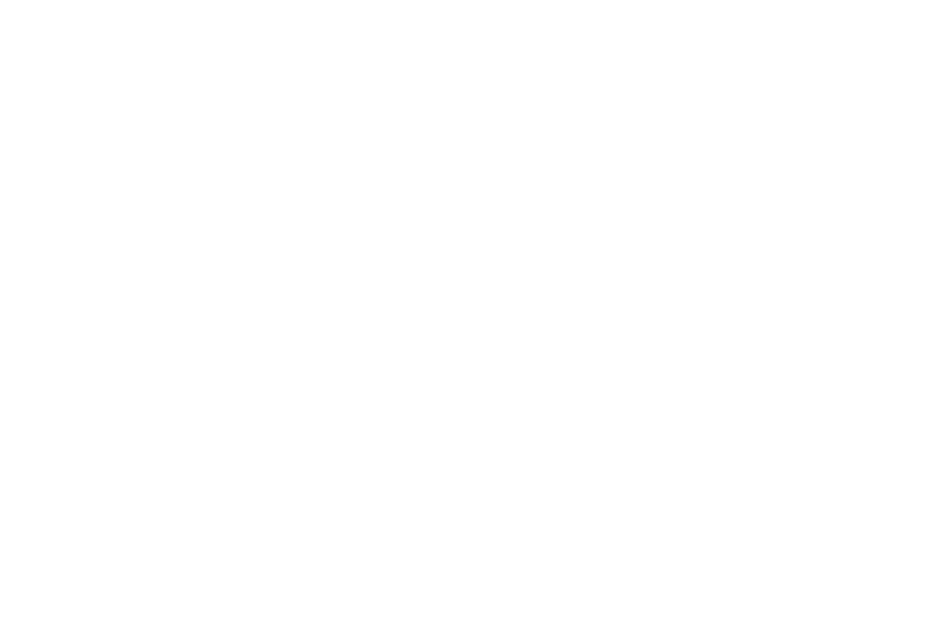}}\\[1ex]
\includegraphics[width=18.5cm]{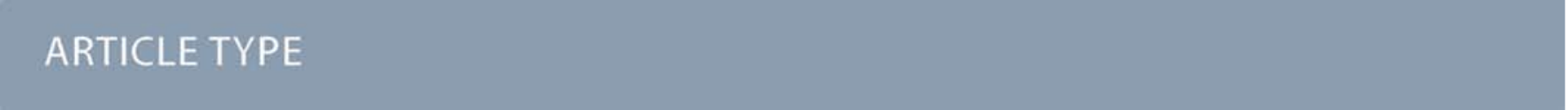}}\par
\vspace{1em}
\sffamily
\begin{tabular}{m{4.5cm} p{13.5cm} }

\includegraphics{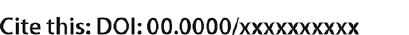} & \noindent\LARGE{\textbf{Crossover from viscous fingering to
fracturing in cohesive wet granular media: a photoporomechanics
study$^\dag$}} \\%Article title goes here instead of the text "This is the title"
\vspace{0.3cm} & \vspace{0.3cm} \\

 & \noindent\large{Yue Meng,\textit{$^{a\ddag}$} Wei Li,\textit{$^{a\P}$} and Ruben Juanes$^{\ast}$\textit{$^{a}$}} \\%Author names go here instead of "Full name", etc.

\includegraphics{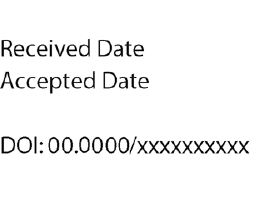} & \noindent\normalsize{We study fluid-induced deformation and fracture of cohesive granular media, and apply photoporomechanics to uncover the underpinning grain-scale mechanics. We fabricate photoelastic spherical particles of diameter $d=2$~mm, and make a monolayer granular pack with tunable intergranular cohesion in a circular Hele-Shaw cell that is initially filled with viscous silicone oil. We inject water into the oil-filled photoelastic granular pack, varying the injection flow rate, defending-fluid viscosity, and intergranular cohesion. We find two different modes of fluid invasion: viscous fingering, and fracturing with leak-off of the injection fluid. We directly visualize the evolving effective stress field through the particles' photoelastic response, and discover a hoop effective stress region behind the water invasion front, where we observe tensile force chains in the circumferential direction. Outside the invasion front, we observe compressive force chains aligning in the radial direction. We conceptualize the system’s behavior by means of a two-phase poroelastic continuum model. The model captures granular pack dilation and compaction with the boundary delineated by the invasion front, which explains the observed distinct alignments of the force chains. Finally, we rationalize the crossover from viscous fingering to fracturing by comparing the competing forces behind the process: viscous force from fluid injection that drives fractures, and intergranular cohesion and friction that resist fractures. } \\%The abstrast goes here instead of the text "The abstract should be..."

\end{tabular}

 \end{@twocolumnfalse} \vspace{0.6cm}

  ]
%%%END OF TITLE, AUTHORS AND ABSTRACT%%%

%%%FONT SETUP - please do not change any commands within this section
\renewcommand*\rmdefault{bch}\normalfont\upshape
\rmfamily
\section*{}
\vspace{-1cm}

%%%FOOTNOTES%%%

\footnotetext{\textit{$^{a}$~Massachusetts Institute of Technology, 77 Massachusetts Avenue, Cambridge, MA, USA; E-mail: juanes@mit.edu}}

%Please use \dag to cite the ESI in the main text of the article.
%If you article does not have ESI please remove the the \dag symbol from the title and the footnotetext below.
\footnotetext{\dag~Electronic Supplementary Information (ESI) available: 2 movie files and a text document with captions for ESI files. See DOI: 10.1039/cXsm00000x/}

\footnotetext{\ddag~Present address: Princeton University, Guyot Hall, Princeton, NJ, USA.}
\footnotetext{\P~Present address: Stony Brook University, Stony Brook, NY, USA. }
%\ddag, \textsection, and \P. Please place the appropriate symbol next to the author's name and include a \texttt{\textbackslash footnotetext} entry in the the correct place in the list.}

%%%END OF FOOTNOTES%%%

%%%MAIN TEXT%%%%

\section{Introduction}
Multiphase flow through granular and porous materials exhibits complex behavior, the understanding of which is critical in many natural and industrial processes, and the design of climate-change mitigation strategies. Examples include infiltration of water into the vadose zone \citep{hill1972wetting}, methane migration in lake sediments \citep{scandella2016ephemerality}, hillslope infiltration and erosion after forest fires \citep{mataix2013soil}, growth and deformation of cells and tissues \citep{charras2005non}, shale gas production \citep{patzek2013gas}, and geological carbon dioxide storage \citep{szulczewski2012lifetime}. This complexity is linked to the interplay between multiphase flow and granular mechanics, which controls the morphological patterns, evolution, and function of a wide range of systems \citep{juanes2020multiphase}. In many granular-fluid systems, the strong coupling among viscous, capillary, and frictional forces leads to a wide range of patterns, including desiccation cracks \citep{groismankaplan94,shinsantamarina11}, labyrinth structures \citep{sandnes2007labyrinth}, granular fingers \citep{cheng2008towards, huang2012granular, zhang2013coupled}, corals, and stick-slip bubbles \citep{sandnes2011patterns}. In the context of interfacial flows, fracture patterns have been observed in loose systems---such as particle rafts as a result of surfactant spreading \citep{peco2017influence,vella2004elasticity}---as well as dense systems---such as colloidal suspensions as a result of drying \citep{dufresne2003flow,goehring2013plasticity}.

While fracturing during gas invasion in fluid-saturated media has been studied extensively in experiments \citep{dufresne2003flow,vella2004elasticity,shin2010fluid,holtzman2012capillary,huang2012granular,goehring2013plasticity,peco2017influence,campbell2017gas,sun2019grain} and simulations \citep{jain2009preferential,holtzman2010crossover,zhang2013coupled,carrier2012numerical,lecampion2015simultaneous,mikelic2015phase,santillan2018phase,meng2020jamming,carrillo2021capillary}, the underlying grain-scale mechanisms behind the morphodynamics and rheologies exhibited by deformable granular media remain poorly understood. To tackle this challenge, \citet{yue22frac} adopt a recently developed experimental technique, \emph{photoporomechanics} \citep{weiLi2021}, to directly visualize the evolving effective stress field in a  fluid-filled cohesionless granular medium during fluid-induced fracturing. The effective stress field exhibits a surprising and heretofore unrecognized phenomenon: behind the propagating fracture tips, an \emph{effective stress shadow}, where the intergranular stress is low and the granular pack exhibits undrained behavior, emerges and evolves as fractures propagate.

Here we aim to extend our previous work \citep{yue22frac} to cohesive granular media. The mechanical and fracture properties of cohesive granular media are of interest for many applications, including powder aggregation \citep{pietsch1969tensile,kendall2001adhesion}, stimulation of hydrocarbon-bearing rock strata for oil and gas production \citep{economides1989reservoir}, preconditioning and cave inducement in mining \citep{van2002hydraulic,jeffrey2013monitoring}, and remediation of contaminated soil \citep{murdoch2002mechanical}. Similar hydraulic fractures manifest naturally at the geological scale, such as magma transport through dikes \citep{spence1987buoyancy,lister1991fluid,rubin1995propagation,roper2005buoyancy,roper2007buoyancy} and crack propagation at glacier beds \citep{tsai2010model,lai2020vulnerability}. Following the early work on cemented aggregates \citep{dvorkin1991effect,dvorkin1994effective,dvorkin1999elastic}, \citet{hemmerle2016cohesive} created a well-defined cohesive granular medium with tunable elasticity by mixing glass beads with curable polymer. Due to the huge stiffness contrast between polymer bridges (kPa--MPa) and glass beads (GPa), the mechanical response of the material is dominated by the deformation of the bridges rather than the deformation of the beads \citep{schmeink2017fracture,hemmerle2021measuring}. There is limited experimental study on weakly sintered or cemented materials \citep{affes2012tensile} with bonds that are of comparable stiffness with that of the grains.  

In this paper, we study fracturing in cohesive wet granular media and extract the evolving effective stress field via photoporomechanics. By mixing photoelastic particles with curable polymer of the same stiffness, we create a well-defined cohesive granular medium with tunable tensile strength. We uncover two modes of water invasion under different injection flow rate, defending fluid viscosity, and tensile strength of the granular pack: viscous fingering, and fracturing with leak-off of the injection fluid. Behind the invasion front, the granular pack is dilated with tensile effective stress in the circumferential direction, while ahead of the invasion front the granular pack is compacted with compressive effective stress in the radial direction. We develop a two-phase poroelastic continuum model to explain the observed distinct force-chain alignments. Finally, we conclude that the competition of intergranular cohesion and friction against viscous force dictates the crossover from viscous fingering to fracturing regime.

%% Material & experimental setup goes here (Fig.1&2)
\section{Materials and Methods}
\begin{figure}
 \centering
 \noindent\includegraphics[height=4cm]{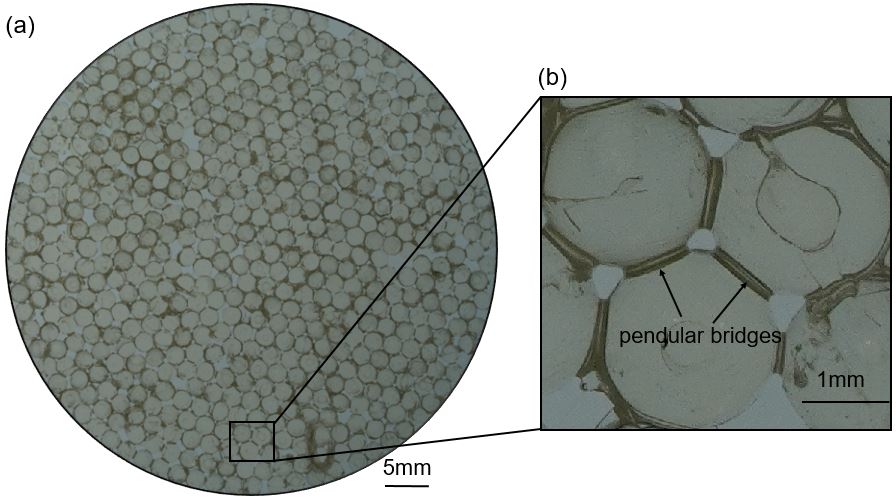}
 \caption{(a) Cohesive photoelastic granular pack at a polymer content $C=4.9\%$. (b) Close-up view of three particles connected by polymer bridges in the form of pendular rings.}
 \label{Fig1_Charac}
\end{figure}

Following the fabrication process in \citep{weiLi2021}, we produce photoelastic spherical particles with diameter $d=1.98$~mm (with $3.5$\% standard deviation) and bulk modulus $K_s=1.6$~MPa. The fabrication process is similar to ``squeeze casting"
for metals, but for polyurethane in our case. The process produces soft polyurethane spheres with
an amber color. To test their sliding frictional properties, we build a shear box apparatus as follows. We prepare a thin acrylic plate in the size of 6~cm x 6~cm x 1.6~mm and punch 11 x 11 holes with diameter 2~mm into it. We squeeze particles into the holes, making sure they are integrated into the plate and can not roll against it. The bottom surface for the sliding test is either made of glass or cured from polyurethane. We then put a confining weight on the top of the acrylic plate, which varies from 2~N to 8~N. We immerse particles in the silicone oil, attach the side surface of the acrylic plate to a spring scale, and drag the plate at a constant velocity 1~mm/s. The spring scale measures the frictional force occurring between the particles and the bottom surface. After dividing it by the confining weight, we obtain the friction coefficient. When particles are immersed in the silicone oil, the friction coefficient between particles is $\mu_p=0.2\pm 0.06$, and the friction coefficient between the particle and the glass plate is $\mu_w=0.05\pm 0.02$. To prepare a cohesive granular pack, we mix a total mass $M_g$ of cured photoelastic particles with a mass $M_{l}$ of uncured, liquid-form polyurethane. We set $M_g=9.2$~g to generate granular packs at a fixed initial two-dimensional packing density, $0.83$, which is close to the random close packing density. We cast the solid-liquid mixture into a monolayer of granular pack inside a circular Hele-Shaw cell. The added polyurethane is imbibed directly into the granular pack and forms polymer bridges between particles that solidify once cured. Before the experiments, we peel the cured monolayer of particles out of the Hele-Shaw cell, eliminating bonds between particles and plates. We define the polymer content $C$ as the mass of added polyurethane divided by the mass of particles, which determines the size of polymer bridges and thus the tensile strength of interparticle bonds. Figure \ref{Fig1_Charac} shows a monolayer of cohesive photoelastic granular pack at a polymer content $C=4.9\%$, above which pendular bridges begin to merge and form clusters \citep{hemmerle2016cohesive}.

\begin{figure}
 \centering
 \noindent\includegraphics[height=4cm]{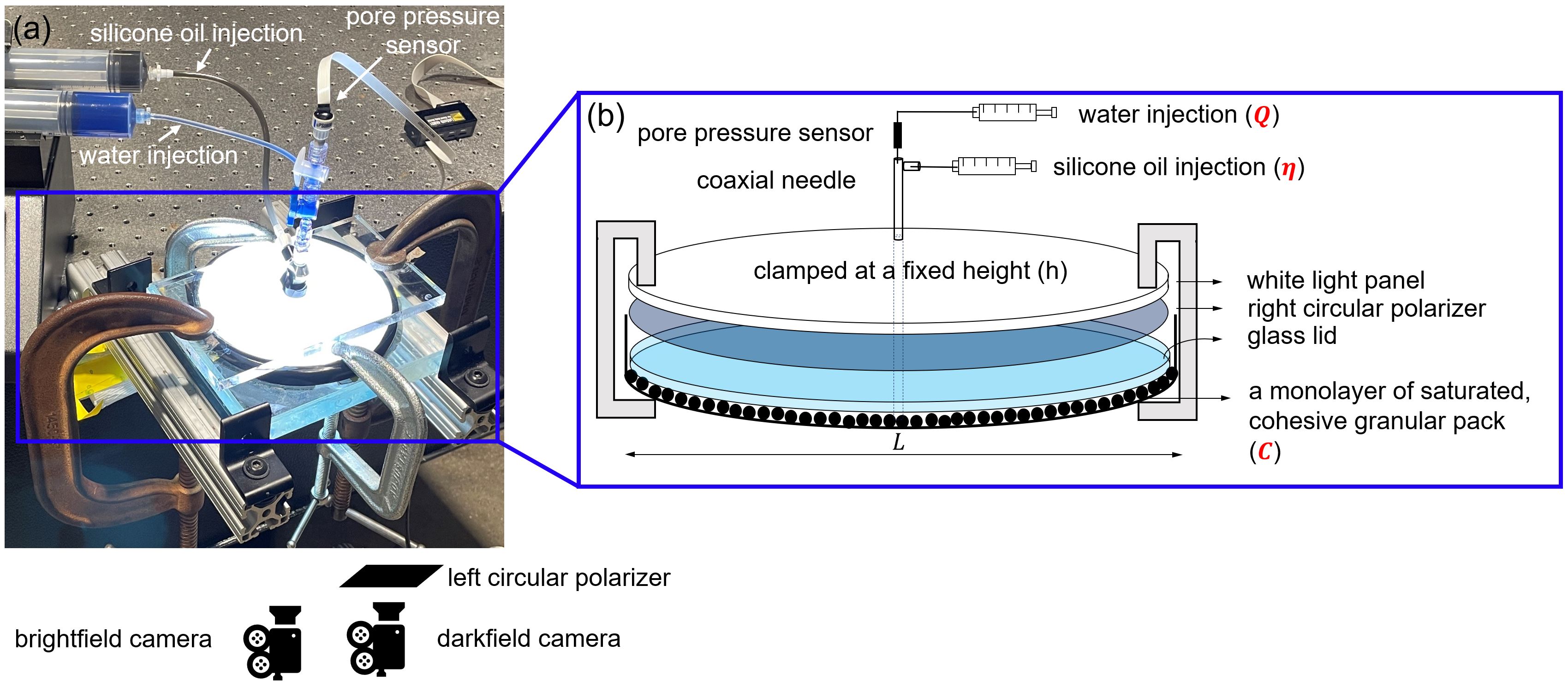}
 \caption[Experimental setup to study fracturing in cohesive photoelastic granular media. (a) A monolayer of photoelastic particles (diameter $d$, polymer content $C$) is confined in a circular Hele-Shaw cell. A white light panel, right and left circular polarizers form a darkfield circular polariscope.  Brightfield and darkfield videos are captured by cameras placed underneath the cell. (b) A close-up of the circular Hele-Shaw cell.]{Experimental setup to study fracturing in cohesive photoelastic granular media. (a) A monolayer of photoelastic particles (diameter $d$, polymer content $C$) is confined in a circular Hele-Shaw cell. The cell is uniformly clamped at a fixed height, $h$. Before the fracturing experiment, silicone oil (viscosity $\eta$) is slowly injected at the center of the cell with a coaxial needle to saturate the granular pack. After saturation, water is injected at a fixed flow rate $Q$ with the injection pressure monitored by a pore-pressure sensor. A white-light panel, right and left circular polarizers form a darkfield circular polariscope.  Brightfield and darkfield videos are captured by cameras placed underneath the cell. (b) Schematic of the circular Hele-Shaw cell (internal diameter $L$). A light panel, a polarizer, and a glass disk rest on top of the monolayer of photoelastic particles. The disk is slightly smaller than the cell to allow the fluids (but not particles) to leave the cell.}
 \label{Fig2_SetUp}
\end{figure}

We inject water into a monolayer of cohesive photoelastic particles saturated with silicone oil in a circular Hele-Shaw cell [Fig.~\ref{Fig2_SetUp}]. To observe the photoelastic response of the particles, we construct a darkfield circular polariscope by a white light panel together with left and right circular polarizers \citep{daniels2017photoelastic}. We clamp the Hele-Shaw cell vertically at a fixed height $h=1.92$~mm with four internal spacers to achieve plane-strain conditions throughout the experiments. As the height of internal spacers is smaller than the particle diameter, the top plate applies vertical confinement on the top of the particles. To allow the fluids (but not the particles) to leave the cell, the disk is made slightly smaller than the interior of the cell (inner diameter $L=10.6$~cm), resulting in a thin gap around the edge of the cell. A coaxial needle is inserted at the center of the granular pack for saturation, fluid injection and pore pressure measurement. We adopt a dual-camera system to record brightfield (camera A) and darkfield (camera B) videos. As water is injected into the cohesive granular pack, viscous forces from fluid injection promote the development of fractures, while intergranular cohesion within the granular pack resists it. To study these competing forces during the fluid invasion process, we vary the injection flow rate $Q$ from 5~mL/min to 220~mL/min, the silicone oil viscosity $\eta$ from 30~kcSt, 100~kcSt, to 300~kcSt, and the polymer content $C$ from $0\%$ to $4.6\%$ to stay in the pendular regime.

%% Interpretation of Fig.3: the representative fracturing experiment, B&D field screenshots.
\section{Representative Experiments}\label{sec_exp}
In this section, we present two representative experiments with $Q=100$~mL/min, $\eta=300$~kcSt, $C=4.4\%$ for viscous fingering, and $C=1.2\%$ for fracturing with leak-off. For the dye color of the injected water, we need one that both visualizes the water invasion front in brightfield images and does not interfere with the photoelastic response in darkfield images. As a result, we dye the defending oil in black, and the invading water in light blue. By tracking the region in light blue color, we could easily identify the invading phase from brightfield images. To confirm this, we refer to the supplemental videos on the fluid morphology and the effective stress evolution for the two experiments \citep{supplement}.

We differentiate between viscous fingering and fracturing with leak-off regimes from the brightfield images. When water invades into the granular pack in viscous fingering patterns without noticeable relative motion between particles, the experiment is classified as viscous fingering (Figure~3). When the injected water creates open channels with ensuing invasion into the pores, then the experiment is classified as fracturing with leak-off (Figure~4). The darkfield images in Figure~4 also confirm the formation of fractures where intergranular bonds exhibit strong photoelastic response and are torn apart under tension.

\subsection{Viscous Fingering}
\begin{figure*}
 \centering
 \noindent\includegraphics[width=\textwidth]{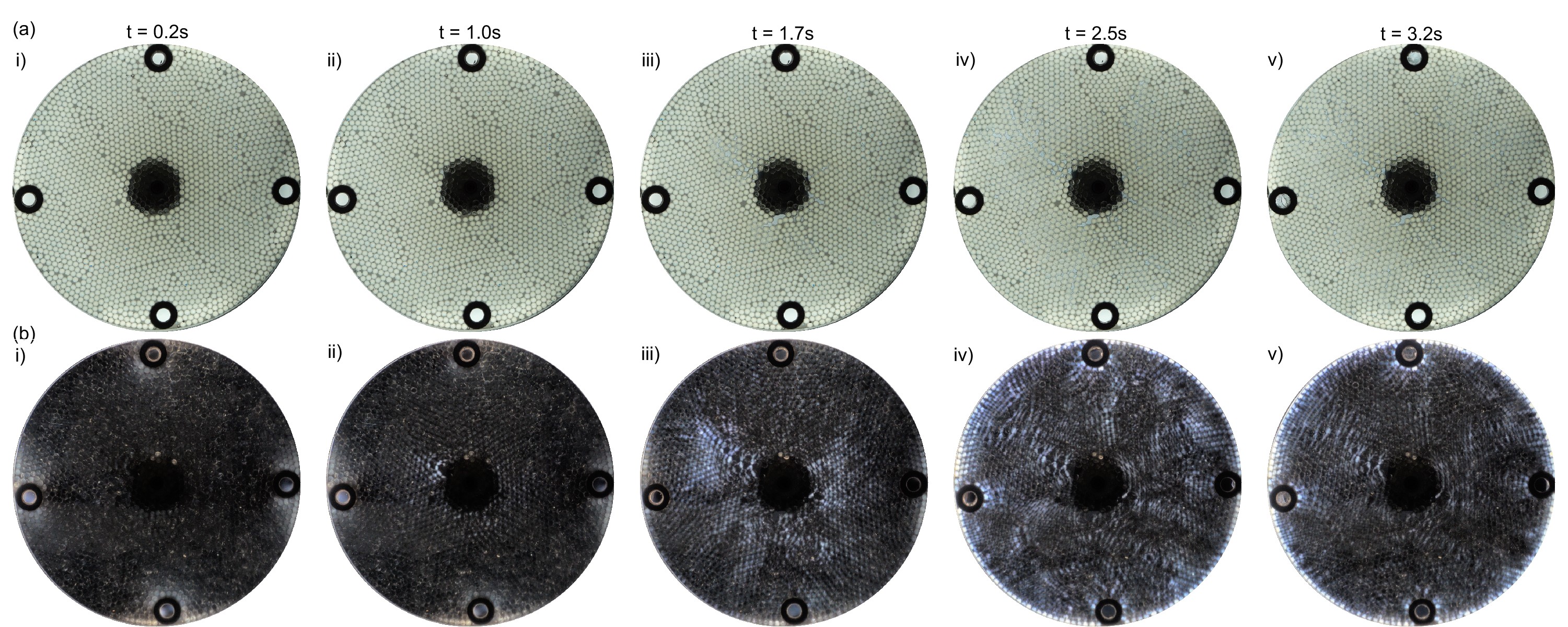}
 \caption[For the viscous fingering experiment with $Q=100$~mL/min, $\eta=300$~kcSt and $C=4.4\%$, a sequence of snapshots shows the time evolution of (a) interface morphology from brightfield images, and (b) effective stress field from darkfield images.]{For the viscous fingering experiment with $Q=100$~mL/min, $\eta=300$~kcSt and $C=4.4\%$, a sequence of snapshots shows the time evolution of (a) interface morphology from brightfield images, and (b) effective stress field from darkfield images. See supplemental video 1 for the evolution of the morphology and effective stress field in this regime.}
 \label{Fig3a_BDEvol}
 \end{figure*}

In Fig.~\ref{Fig3a_BDEvol}, we show a sequence of snapshots for the viscous fingering experiment. We analyze the time evolution of the water--oil interface morphology from brightfield images, and the interparticle stresses of the granular pack from darkfield images. When particles have been strongly cemented initially, the injection pressure is insufficient to overcome the tensile strength of the intergranular bonds. In this regime, we observe patterns of viscous fingers without any significant relative motion between particles [Fig.~\ref{Fig3a_BDEvol}(a)], and the intergranular bonds at finger tips endure tension without breakage [Fig.~\ref{Fig3a_BDEvol}(b)]. 

\subsection{Fracturing with Leak-off}
\begin{figure*}
 \centering
 \noindent\includegraphics[width=\textwidth]{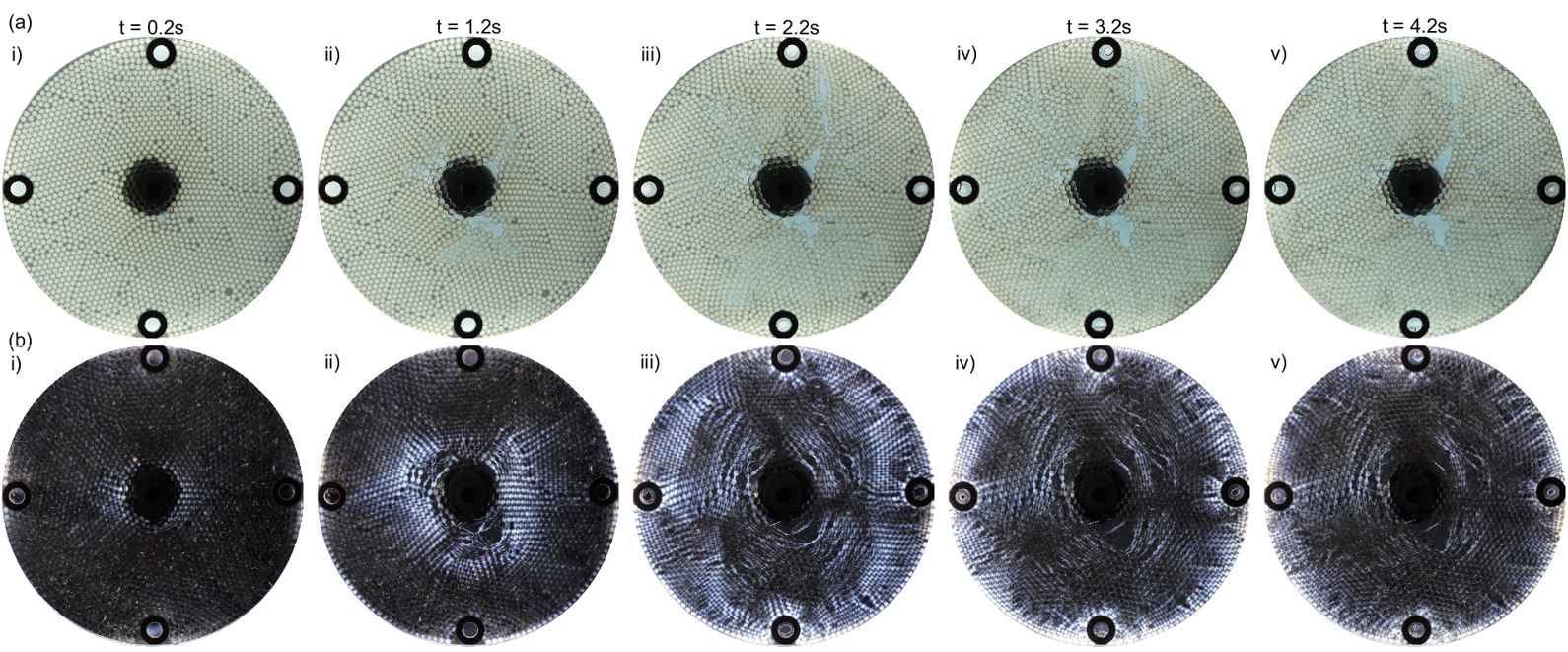}
 \caption[For the fracturing experiment with $Q=100$~mL/min, $\eta=300$~kcSt and $C=1.2\%$, a sequence of snapshots shows the time evolution of (a) interface morphology from brightfield images, and (b) effective stress field from darkfield images.]{For the fracturing experiment with $Q=100$~mL/min, $\eta=300$~kcSt and $C=1.2\%$, a sequence of snapshots shows the time evolution of (a) interface morphology from brightfield images, and (b) effective stress field from darkfield images. See supplemental video 2 for the evolution of the morphology and effective stress field in this regime.}
 \label{Fig3_BDEvol}
\end{figure*}

In Fig.~\ref{Fig3_BDEvol}, we show a sequence of snapshots for the fracturing experiment. The time evolution of the injection pressure $P_{\text{inj}}$ is plotted in Fig.~\ref{Fig4_ModPressS}(a), which also indicates the times of the snapshots in Fig.~\ref{Fig3_BDEvol}. When particles have been densely packed initially, water firstly invades into the cohesive granular pack by expanding a small cavity around the injection port, with $P_{\text{inj}}$ ramping up during this period. The pressure keeps building up until it becomes sufficient to overcome the tensile strength of the interparticle bonds; the point at which fractures emerge [in between Fig.~\ref{Fig3_BDEvol}(i),(ii)]. As injection continues, much of the injected water leaks off into the permeable granular media as the fractures propagate [Fig.~\ref{Fig3_BDEvol}(ii)$\sim$(v)]. In this period of fracturing with leak-off, $P_{\text{inj}}$ slightly drops from its peak value and reaches a plateau afterwards [Fig.~\ref{Fig4_ModPressS}(a)]. In this regime, the effective stress field exhibits a surprising phenomenon: behind the water invasion front, a \emph{hoop effective stress region}, where we observe tensile force chains in the circumferential direction, emerges and evolves as invasion front propagates. Ahead of the invasion front, we observe compressive effective stress in the radial direction [Fig.~\ref{Fig3_BDEvol}(b)]. The phenomena regarding the effective stress (e.g. tensile hoop stress near the injection port) has been predicted by continuum theories, such as cavity expansion models for single-phase flow \citep{auton2018arteries,auton2019large}, and tip asymptotics in fracture mechanics (Sections 2 and 3 in \citep{detournay16}). However, there is a lack of understanding of the effective stress evolution in a two-phase immiscible flow system, and our experiments visualize it for the first time.

%% Discussion on the continuum model goes here
 \section{Two-Phase Poroelastic Continuum Model} 
 We model the immiscible injection of water into a cohesive granular pack saturated with silicone oil. To rationalize the experimental findings in section \ref{sec_exp}, we develop a two-phase poroelastic continuum model focusing on the fracturing with leak-off regime. The wetting phase is the defending oil, and the nonwetting phase is the invading water. Under the experimental conditions explored, the modified capillary number \citep{holtzman2012capillary} $\text{Ca}^*=\eta QL/(\gamma hd^2)\sim10^6$, which indicates that viscous forces outweigh capillary forces so we can safely neglect capillary effects. In the following, we present the extension of Biot's theory \citep{biot41} to two-phase flow \citep{jha2014coupled,bjornaraa2016vertically}. In our model, we assume radial symmetry and small deformations. 
 
\subsection{Fluid Flow Equations}
For the two-phase immiscible flow system, the conservation of fluid mass can be written as follows:
\begin{equation}
    \frac{\partial(\phi\rho_\alpha S_\alpha)}{\partial t}+\nabla\cdot(\rho_\alpha\phi S_\alpha \mathbf{v_\alpha})=0,
    \label{eqn_FldCont}
\end{equation}
where $\phi$ is the porosity and $\rho_\alpha$ and $S_\alpha$ are the density and saturation of the fluid phase $\alpha$ (water $w$ or oil $o$), respectively. The phase velocity $\mathbf{v_\alpha}$ is related to the Darcy flux $\mathbf{q_\alpha}$ in a deforming medium by the following relation:
\begin{equation}
    \mathbf{q_\alpha}=\phi S_\alpha(\mathbf{v_\alpha}-\mathbf{v_s})=-\frac{k_0}{\eta_\alpha}k_{r\alpha}(\nabla p_\alpha-\rho_\alpha\mathbf{g})
    \label{eqn_DarcyFlow}
\end{equation}
where $\mathbf{v_s}$ is the velocity of the solid skeleton, $k_0$ is the intrinsic permeability of the granular pack, $\mathbf{g}$ is the gravity vector, and $\eta_\alpha$, $k_{r\alpha}=k_{r\alpha}(S_\alpha)$ and $p_\alpha$ are the dynamic viscosity, relative permeability, and fluid pressure for phase $\alpha$, respectively. Since capillary pressure is negligible here, $p_c=p_w-p_o=0$, the two phases have the same fluid pressure $p$. The relative permeability functions are given as Corey-type power law functions \citep{brooks1964hydraulic}:
\begin{equation}
    k_{rw}=\left(\frac{S_w-S_{wc}}{1-S_{wc}}\right)^{a_w}, \quad k_{ro}=\left(1-\frac{S_w}{1-S_{ro}}\right)^{a_o},
\end{equation}
where the fitting parameters are the critical water saturation for water to flow, $S_{wc}=0.2$, the residual oil saturation, $S_{ro}=0.2$, and the exponents $a_w=2$ and $a_o=5$. 

Considering the mass conservation of the solid phase:
\begin{equation}
    \frac{\partial[\rho_s(1-\phi)]}{\partial t}+\nabla\cdot[\rho_s(1-\phi)\mathbf{v_s}]=0,
    \label{eqn_SolCont}
\end{equation}
where $\rho_s$ is the density of the solid constituents of the porous medium. Assuming isothermal conditions and using the equation of state for the solid, the following expression for the change in porosity is obtained \citep{lewis1998finite}:
\begin{equation}
    \frac{d\phi}{dt}=(b-\phi)\left(c_s\frac{dp}{dt}+\nabla\cdot\mathbf{v_s}\right)
    \label{eqn_poroLaw}
\end{equation}
where $b$ is the Biot coefficient of the saturated porous medium, and $c_s$ is the compressibility of the solid phase. We use equations \eqref{eqn_DarcyFlow}, \eqref{eqn_SolCont}, and \eqref{eqn_poroLaw} to expand equation \eqref{eqn_FldCont} as follows:
\begin{equation}%\tag{Eqn. 1}
\phi\frac{\partial S_\alpha}{\partial t}+S_\alpha\left(b\frac{\partial \epsilon_{kk}}{\partial t}+\frac{1}{M}\frac{\partial p}{\partial t}\right)+\nabla\cdot \mathbf{q}_\alpha=0,
    \label{eqn_TPfldcont}
\end{equation}
where $\epsilon_{kk}$ is the volumetric strain of the solid phase. The Biot modulus of the saturated granular pack, $M$, is related to fluid and rock properties as $\frac{1}{M}=\phi c_f+(b-\phi)c_s$ \citep{coussy1995mechanics}. Adding equation \eqref{eqn_TPfldcont} for water and oil phases, and imposing that $S_o+S_w\equiv1$ for the saturated granular pack, we obtain the pressure diffusion equation:
\begin{equation}
b\frac{\partial\epsilon_{kk}}{\partial t}+\frac{1}{M}\frac{\partial p}{\partial t}+\nabla\cdot\mathbf{q}_t=0,
    \label{eqn_pressDiff}
\end{equation}
where $\mathbf{q}_t$ is the total Darcy flux for water and oil phases, $\mathbf{q}_t=\mathbf{q}_w+\mathbf{q}_o$. 

\subsection{Geomechanical Equations} 
Under quasi-static conditions, the balance of linear momentum of the solid-fluid system states that:
\begin{equation}
\nabla\cdot\boldsymbol{\sigma}+\rho_b\mathbf{g}=\mathbf{0},
    \label{eqn_forBal}
\end{equation}
where $\boldsymbol{\sigma}$ is the Cauchy total stress tensor, and $\rho_b=(1-\phi)\rho_s+\phi\sum\limits_{\alpha}\rho_\alpha S_\alpha$, is the bulk density for the solid-fluid system. For axisymmetric deformation in plane-strain condition, the force balance equation becomes:
\begin{equation}
\frac{\partial\sigma_{rr}}{\partial r}+\frac{\sigma_{rr}-\sigma_{\theta\theta}}{r}=0.
    \label{eqn_forBal_r}
\end{equation}

Following \citep{coussy1995mechanics}, the poroelasticity equation states that
\begin{equation}
\boldsymbol{\sigma}=\boldsymbol{\sigma'}-bp\boldsymbol{\mathbb{I}},
    \label{eqn_poroElast}
\end{equation}
where $\boldsymbol{\mathbb{I}}$ represents the identity matrix. Terzaghi's effective stress tensor $\boldsymbol{\sigma'}$ is the portion of the stress supported through deformation of the solid skeleton, and where we adopt the convention of tension being positive. We adopt isotropic linear elastic theory for the granular pack; the constitutive equation for stress--strain is:
\begin{equation}
\boldsymbol{\sigma'}=\frac{3K\nu}{1+\nu}\epsilon_{kk}\boldsymbol{\mathbb{I}}+\frac{3K(1-2\nu)}{1+\nu}\boldsymbol{\epsilon},
    \label{eqn_linElas}
\end{equation}
where $K$, $\nu$ are the drained bulk modulus, and the drained Poisson ratio of the granular pack, respectively. The strain tensor is defined as $\boldsymbol{\epsilon}=\frac{1}{2}[\nabla\mathbf{u}+(\nabla\mathbf{u})^T]$, where $\mathbf{u}$ is the displacement vector. For the axisymmetric deformation in plane-strain condition, the strains are written as:
\begin{equation}
\epsilon_{rr}=\frac{\partial u_r}{\partial r},\quad \epsilon_{\theta\theta}=\frac{u_r}{r}, \quad\epsilon_{zz}=0.
    \label{eqn_strains}
\end{equation}
Using equations \eqref{eqn_poroElast}, \eqref{eqn_linElas} and \eqref{eqn_strains}, the force balance equation \eqref{eqn_forBal_r} can be expressed as a function of radial displacement $u_r(r,t)$ and pore pressure $p(r,t)$. 
\subsection{Summary of Governing Equations}
The model has three governing equations, two derived from conservation of fluid mass [Eqn.~\eqref{eqn_pressDiff} for the water--oil fluid mixture and Eqn.~\eqref{eqn_TPfldcont} for the water phase] and one derived from conservation of linear momentum [Eqn.~\eqref{eqn_forBal_r}]. The model solves the time evolution of three unknowns: (1) pore pressure field $p(r,t)$; (2) water saturation field $S_w(r,t)$; and (3) radial displacement field $u_r(r,t)$ of the cohesive granular pack, all of which are assumed to be radially symmetric. The governing equations are summarized and written in radial coordinates as follows:
\begin{align}
b\frac{\partial\epsilon_{kk}}{\partial t}+\frac{1}{M}\frac{\partial p}{\partial t}-k_0\frac{\partial}{\partial r}\left(\left(\frac{k_{rw}}{\eta_w}+\frac{k_{ro}}{\eta_o}\right)\frac{\partial p}{\partial r}\right)-\frac{k_0}{r}\left(\frac{k_{rw}}{\eta_w}+\frac{k_{ro}}{\eta_o}\right)\frac{\partial p}{\partial r}&=0,\label{eqn_govTP_press}\\
\phi\frac{\partial S_w}{\partial t}+S_w\left(b\frac{\partial \epsilon_{kk}}{\partial t}+\frac{1}{M}\frac{\partial p}{\partial t}\right)-\frac{k_0}{\eta_w}\frac{\partial}{\partial r}\left(k_{rw}\frac{\partial p}{\partial r}\right)-\frac{k_0}{r}\frac{k_{rw}}{\eta_w}\frac{\partial p}{\partial r}&=0,\label{eqn_govSP_waterPress}\\
\frac{\partial\sigma_{rr}}{\partial r}+\frac{\sigma_{rr}-\sigma_{\theta\theta}}{r}&=0\label{eqn_gov_force}.
\end{align}
For the axisymmetric deformation in plane-strain condition, the stresses and strains are written in radial coordinates as:
\begin{align}
\sigma_{rr}&=\frac{3K\nu}{1+\nu}\epsilon_{kk}+\frac{3K(1-2\nu)}{1+\nu}\epsilon_{rr}-bp,\\
\sigma_{\theta\theta}&=\frac{3K\nu}{1+\nu}\epsilon_{kk}+\frac{3K(1-2\nu)}{1+\nu}\epsilon_{\theta\theta}-bp,\\
\epsilon_{rr}&=\frac{\partial u_r}{\partial r},\quad \epsilon_{\theta\theta}=\frac{u_r}{r}, \quad\epsilon_{zz}=0,\\
\epsilon_{kk}&=\epsilon_{rr}+\epsilon_{\theta\theta}+\epsilon_{zz}.
\end{align}

We initialize the model by specifying $u_r(r,0)=p(r,0)=S_w(r,0)=0$. As for the boundary conditions, the inner boundary is free to move, subject to injection pressure as the total stress: 
\begin{equation}
\sigma_{rr}(r_0,t)=-p(r_0,t)=-P_{\text{inj}}(t),
    \label{eqn_bc1Stress}
\end{equation}
where $r=r_0$ is the radius of the injection port, and $P_{\text{inj}}(t)$ is the injection pressure at time $t$. At the injection port, the total Darcy flux is the same as the Darcy flux of water. Since the injection system is composed of plastic syringe and tubing, we take the system compressibility into account for the inner flow boundary condition:
\begin{equation}
q_t(r_0,t)=q_w(r_0,t)=\frac{Q-c_{\text{sys}}V_{\text{sys}}\frac{\partial P_{\text{inj}}(t)}{\partial t}}{2\pi r_0h},
    \label{eqn_bc1Flow}
\end{equation}
where $Q$ is the injection flow rate, and $c_{\text{sys}}$ and $V_{\text{sys}}$ are the compressibility and volume of the injection system, respectively. At the outer boundary $r=R$, the pressure is atmospheric, and particles have zero radial displacement:
\begin{equation}
p(R,t)=u_r(R,t)=0.
    \label{eqn_bc2}
\end{equation}

%% Dimensional Analysis
We now summarize the model in dimensionless form, denoting dimensionless quantities with a tilde. We adopt characteristic scales for length, time, stress/pressure, viscosity and permeability, non-dimensionalizing the governing equations via the scaling
\begin{equation}
\begin{aligned}
    \tilde{r}=\frac{r}{R},\quad \tilde{u}_r=\frac{u_r}{R},\quad \tilde{t}=\frac{t}{T_{\text{pe}}},\quad \tilde{p}=\frac{p}{M},&\quad \tilde{\eta}_{\alpha}=\frac{\eta_\alpha}{\eta_o}, \quad \tilde{k}_\alpha=\frac{k_0k_{r\alpha}}{k_0},\quad \\ \tilde{\sigma}_{rr}=\frac{\sigma_{rr}}{K},\quad  \tilde{\sigma}_{\theta\theta}=\frac{\sigma_{\theta\theta}}{K},\quad
    &\tilde{\sigma}_{rr}'=\frac{\sigma_{rr}'}{K},\quad \tilde{\sigma}_{\theta\theta}'=\frac{\sigma_{\theta\theta}'}{K},
\end{aligned}
\end{equation}
where $T_{\text{pe}}=\frac{\eta_oR^2}{k_0M}$ is the characteristic poroelastic timescale. We can then rewrite Eqn.~\eqref{eqn_govTP_press},\eqref{eqn_govSP_waterPress},\eqref{eqn_gov_force} in dimensionless form,
\begin{align}
b\frac{\partial\epsilon_{kk}}{\partial \tilde{t}}+\frac{\partial \tilde{p}}{\partial \tilde{t}}-\frac{\partial}{\partial \tilde{r}}\left(\left(\frac{k_{rw}}{\tilde{\eta}_w}+\frac{k_{ro}}{\tilde{\eta}_o}\right)\frac{\partial \tilde{p}}{\partial \tilde{r}}\right)-\frac{1}{\tilde{r}}\left(\frac{k_{rw}}{\tilde{\eta}_w}+\frac{k_{ro}}{\tilde{\eta}_o}\right)\frac{\partial \tilde{p}}{\partial \tilde{r}}&=0,\label{eqn_govTP_press_DA}\\
\phi\frac{\partial S_w}{\partial \tilde{t}}+S_w\left(b\frac{\partial \epsilon_{kk}}{\partial \tilde{t}}+\frac{\partial \tilde{p}}{\partial \tilde{t}}\right)-\frac{1}{\tilde{\eta}_w}\frac{\partial}{\partial \tilde{r}}\left(k_{rw}\frac{\partial \tilde{p}}{\partial \tilde{r}}\right)-\frac{1}{\tilde{r}}\frac{k_{rw}}{\tilde{\eta}_w}\frac{\partial \tilde{p}}{\partial \tilde{r}}&=0,\label{eqn_govSP_waterPress_DA}\\
\frac{\partial\tilde{\sigma}_{rr}}{\partial \tilde{r}}+\frac{\tilde{\sigma}_{rr}-\tilde{\sigma}_{\theta\theta}}{\tilde{r}}&=0\label{eqn_gov_force_DA}.
\end{align}
where the dimensionless stresses are written in radial coordinates as:
\begin{align}
\tilde{\sigma}_{rr}&=\frac{3\nu}{1+\nu}\epsilon_{kk}+\frac{3(1-2\nu)}{1+\nu}\epsilon_{rr}-\frac{bM}{K}\tilde{p},\\
\tilde{\sigma}_{\theta\theta}&=\frac{3\nu}{1+\nu}\epsilon_{kk}+\frac{3(1-2\nu)}{1+\nu}\epsilon_{\theta\theta}-\frac{bM}{K}\tilde{p}.
\end{align}

We initialize the model by specifying $\tilde{u}_r(\tilde{r},0)=\tilde{p}(\tilde{r},0)=S_w(\tilde{r},0)=0$. The boundary conditions are as follows:
\begin{equation}
\begin{aligned}
\tilde{\sigma}_{rr}(\tilde{r}_0,\tilde{t})&=-\frac{\tilde{P}_{\text{inj}}(\tilde{t})M}{K}, \\
\tilde{q}_t(\tilde{r}_0,\tilde{t})&=\tilde{q}_w(\tilde{r}_0,\tilde{t})=\frac{\tilde{Q}R}{\tilde{r}_0h}-\frac{c_\text{sys}V_\text{sys}M}{2\pi r_0kh}, \\
\tilde{p}(1,\tilde{t})&=\tilde{u}_r(1,\tilde{t})=0.
\end{aligned}
\end{equation}
where $\tilde{P}_{\text{inj}}(t)=\frac{P_{\text{inj}}(t)}{M}$, $\tilde{Q}=\frac{\eta_o Q}{2\pi k_0MR}$. Both of these
quantities compare the viscous pressure due to injection with the Biot modulus of the granular pack. 

\subsection{Model Parameters}
The four poroelastic constants in the model are the drained bulk modulus $K$, the drained Poisson ratio $\nu$, the Biot coefficient $b$, and the Biot modulus $M$ of the granular pack. We obtained the drained and undrained bulk modulus $K$, $K_u$ from a separate consolidation experiment \citep{weiLi2021}. We calculate the Biot coefficient from the relationship $b=1-\frac{K}{K_s}$ \citep{coussy1995mechanics}, and then obtain the Biot modulus via  $M=\frac{K_u-K}{b^2}$ \citep{wang2000theory}. To obtain the drained Poisson ratio of the granular pack, we build a discrete element model and conduct a biaxial test \citep{itasca-pfc2d}. The permeability of the granular pack, $k$, is measured experimentally during the initial oil saturation process. A summary of the modeling parameters is given in Table~\ref{Tab:ModParam}.    
 \begin{table}
 \caption{Modeling parameters for the two-phase poroelastic continuum model}
 \centering
\begin{tabular*}{0.54\textwidth}{l c l c}
 \hline
  Symbol  & Value & Unit & Variable  \\
 \hline
  $r_0$ & 4&mm& Injection port radius \\
  $R$ &5.3 &cm& Hele-Shaw cell radius\\
  $d$&2&mm&Diameter of the photoelastic particles\\
  $h$&1.92&mm&Height of the Hele-Shaw cell\\
  $Q$&100&mL/min&Water injection flow rate\\
  $c_{\text{sys}}$&6$\times10^{-8}$&Pa$^{-1}$& Injection system compressibility\\
  $V_{\text{sys}}$&30&mL&Injection system volume\\
   $K$ & 200 & kPa & Drained bulk modulus of the pack   \\
   $K_u$ & 1.35 & MPa & Undrained bulk modulus of the pack\\
   $\nu$  & 0.4 & & Drained Poisson ratio of the pack   \\
   $b$  & 0.88 & & Biot coefficient of the pack   \\
   $M$  & 1.49 & MPa & Biot modulus of the pack   \\
  $\eta_w$ & 0.001 & Pa$\cdot$s& Injecting water viscosity    \\
  $\eta_o$ & 291.3 & Pa$\cdot$s& Defending silicone oil viscosity    \\
   $\phi$  & 0.4 & & Porosity of the pack   \\
   $k_0$  & 10$^{-10}$ & m$^2$ & Intrinsic permeability of the pack   \\
 \hline
 %\multicolumn{2}{l}{$^{a}$Footnote text here.}
 \end{tabular*}\label{Tab:ModParam}
 \end{table}
\subsection{Numerical Implementation}
We use a finite volume numerical scheme to solve the three coupled governing equations [Eqns.~\eqref{eqn_govTP_press},~\eqref{eqn_govSP_waterPress},~\eqref{eqn_gov_force}]. We place the pressure and saturation unknowns ($p(r,t),S_w(r,t)$) at volume centers, and displacement unknowns ($u_r(r,t)$) at nodes. We partition the coupled problem and solve two sub-problems sequentially: the coupled flow and mechanics, and the transport of water saturation. We first fix the water saturation, and solve the coupled flow and mechanics equations [Eqns.~\eqref{eqn_govTP_press},~\eqref{eqn_gov_force}] simultaneously using implicit time discretization. Then we solve the water transport equation [Eqn.~\eqref{eqn_govSP_waterPress}] with prescribed pressure and displacement fields. 

\section{Results and Discussion}
 
\subsection{Pore Pressure and Water Saturation}
\begin{figure*}
 \centering
 \noindent\includegraphics[height=4cm]{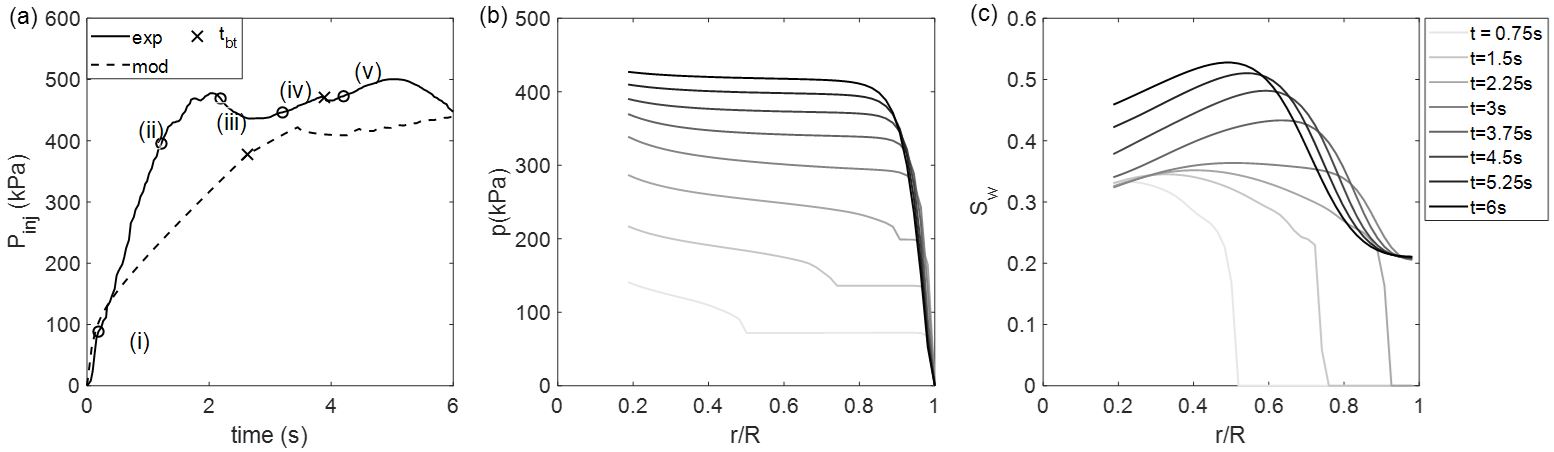}
 \caption[Modeling results for the fracturing experiment with $Q=100$~mL/min, $\eta=300$~kcSt and $C=1.2\%$. (a) Time evolution of the injection pressure $P_{\text{inj}}$. Modeling results of the time evolution of (b) pore pressure $p(r,t)$, and (c) water saturation $S_w(r,t)$.]{Modeling results for the fracturing experiment with $Q=100$~mL/min, $\eta=300$~kcSt and $C=1.2\%$. (a) Time evolution of the injection pressure $P_{\text{inj}}$. The solid curve represents the experimental measurement, and the dashed curve represents the model prediction. The cross markers indicate water breakthrough through the cell edge. The circular markers indicate times for the snapshots shown in Figs.~\ref{Fig3_BDEvol},~\ref{Fig5_ExpModUrt},~\ref{Fig6_ExpModEpsilon},~\ref{Fig8_ExpMod_effStr}: t=0.2s, 1.2s, 2.2s, 3.2s, 4.2s in sequence. Modeling results of the time evolution of (b) pore pressure $p(r,t)$, and (c) water saturation $S_w(r,t)$. We show the solution at 8 times, linearly spaced from t = 0.75s (light gray) to t = 6s (black).}
 \label{Fig4_ModPressS}
\end{figure*}

We compare the experimental and modeling results of the time evolution of the injection pressure $P_{\text{inj}}$ for the case with $Q=100$~mL/min, $\eta=300$~kcSt and $C=1.2\%$ [Fig.~\ref{Fig4_ModPressS}(a)]. By taking the injection system compressibility into account, the model captures the initial $P_{\text{inj}}$ ramp-up measured in the experiment ($t=0\sim0.3$~s). Before $t=3.5$~s, $P_{\text{inj}}$ keeps increasing, with the diffusion of pore pressure [Fig.~\ref{Fig4_ModPressS}(b)] and propagation of water invasion front [Fig.~\ref{Fig4_ModPressS}(c)]. The transient pressure response comes from the compressibility of the granular pack, the timescale of which is $T\sim\frac{\eta'R^2}{k'M}$, where $\eta'$ and $k'$ are the effective viscosity and permeability of the pore fluid: a water-oil mixture. As the pore pressure diffuses to the cell boundary, $P_{\text{inj}}$ approaches its steady state value, $P^{\text{ss}}_{\text{inj}}\sim\frac{\eta'QR}{2\pi k'r_0h}$. 

The cross markers in Fig.~\ref{Fig4_ModPressS}(a) represent the moment when water reaches the cell boundary for the experiment and the model. The breakthrough predicted by the model is faster than that of the experiment by around 1.2s. The observation that the water invasion front propagates faster in the model hints at an overestimation of the Biot modulus $M$; in other words, the model underestimates the granular pack compressibility/storativity. It reveals two underlying model limitations: (1) the storativity in the model, $S=\frac{1}{M}$, is assumed to be a constant without spatiotemporal variations, which in the experiment increases with porosity in the region where the granular pack dilates; and (2) by assuming linear elastic granular packs with small deformations, the model cannot capture the significant increase in storativity arising from the opening of fracture, where the porosity of the granular pack locally increases to 1. 

%% Discussion of the modeling result on water saturation field goes as follows
Solving the water transport equation [Eqn.~\eqref{eqn_govSP_waterPress}], we obtain the time evolution of the water saturation field [Fig.~\ref{Fig4_ModPressS}(c)]. The water invasion front propagates with the injection until its breakthrough at t=2.6s. After breakthrough, the radial profile of water saturation is nonmonotonic, exhibiting an increase of $S_w$ with $r$ and then a decrease. The position of the local maximum of the saturation profile moves towards the center of the cell as time evolves, and the value of the maximum saturation increases with time. This unusual behavior of water saturation contrasts that of fluid-fluid displacement in a rigid porous medium \citep{buckley1942mechanism, blunt2017multiphase} and highlights the strong coupling between fluid flow and medium deformation in our system. 

\subsection{Displacement and Volumetric Strain}
%% Discussion on displacement field goes as follows
To probe into the granular mechanics behind the fracturing experiment in Fig.~\ref{Fig3_BDEvol}, we first measure the internal deformation of the pack via particle tracking, which provides a direct measure of the displacement field. We define a rectangular coordinate system centered at the injection port, where ($x_i,y_i$) is the position of particle $i$ at time $t$ and ($X_i,Y_i$) is its initial position. The displacement of particle $i$ is then $\mathbf{u_i}=(x_i-X_i, y_i-Y_i)$, with magnitude $u_i(t)=\sqrt{(x_i-X_i)^2+(y_i-Y_i)^2}$ and radial component $u_{r,i}(t)=\sqrt{x^2_i+y^2_i}-\sqrt{X^2_i+Y^2_i}$. The deformation is primarily radial because of the axisymmetric boundary conditions, so we focus on $u_r$. 

\begin{figure*}
 \centering
 \noindent\includegraphics[width=\textwidth]{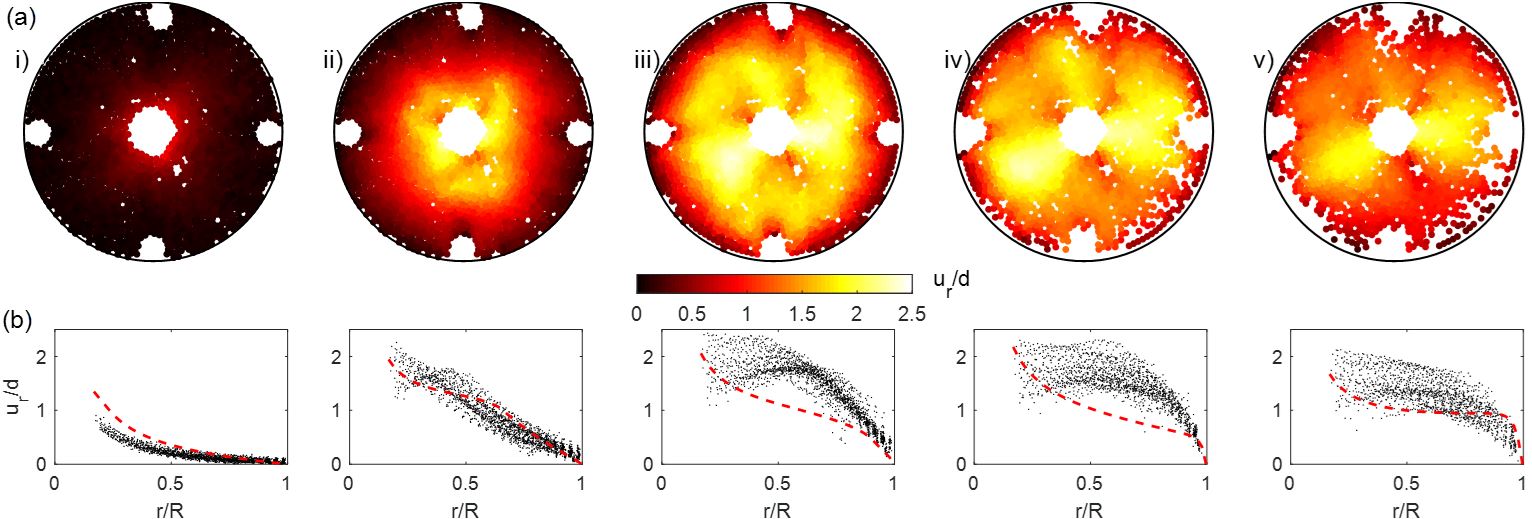}
 \caption{For the fracturing experiment with $Q=100$~mL/min, $\eta=300$~kcSt and $C=1.2\%$, we analyze the sequence of snapshots shown in Fig.~4: i), ii), iii), iv), v) corresponding to t=0.2s, 1.2s, 2.2s, 3.2s, 4.2s, respectively. The sequence of snapshots shows the time evolution of (a) experimental radial displacement field, and (b) radial displacement of the particles (black dots) compared with the continuum model prediction (dashed line).}
 \label{Fig5_ExpModUrt}
\end{figure*}

Figure \ref{Fig5_ExpModUrt} shows a sequence of snapshots of the experimental radial displacement field, corresponding to t=0.2s, 1.2s, 2.2s, 3.2s, 4.2s sequentially. We find that the radial displacement is large near the injection port and fades to zero at the rigid outer edge, with a petal-like mesoscale structure as reported by \citet{macminn2015fluid} and \citet{zhang2011coupled} [Fig.~\ref{Fig5_ExpModUrt}(a)]. The radial displacements of particles are plotted as black dots in Fig.~\ref{Fig5_ExpModUrt}(b), and the red dashed line is the prediction from the continuum model. As $P_{\text{inj}}$ increases from snapshots (i) to (iii), particles move radially outwards. From snapshots (iii) to (v), $P_{\text{inj}}$ reaches a plateau, and particles relax and recover part of the deformation. The model captures the general trends in particle displacement behavior, with the notable exception of the compaction front near $r\sim0.5R$ between times (iii) and (iv). Between this time period, the experimental data shows that particles with $r<0.5R$ are compacted to a similar extent, as evidenced by their similar $u_r$ values, which we refer to as a compaction front. The model underestimates the displacements there due to our assumption of linear elastic behavior: it cannot capture the plasticity-induced compaction front brought by bond breakage and particle rearrangements. As a result, the model fails to capture the compaction front exhibited in the experiment, which we define as the plasticity-induced compaction front. 

%% Discussion on volumetric strain field goes as follows
We use the particle positions to calculate a best-fit local strain field. At time $t$ during the water injection, we compute the closest possible approximation to a local strain tensor $\boldsymbol{\epsilon}$ in the neighborhood of any particle with a sampling radius $r_s=3d$ \citep{falk1998dynamics,macminn2015fluid}. The local strain for the particle is determined by minimizing the mean-square difference $D^2$ between the actual displacements of the neighboring particles relative to the central one and the relative displacements that they would have if they were in a region of uniform strain $\epsilon_{ij}$. That is, we define
\begin{equation}
\begin{aligned}
    D^2(t)=&\sum_n\left(x_n-x_0-(1+\epsilon_{11})(X_n-X_0)-\epsilon_{12}(Y_n-Y_0)\right)^2\\
    &+\left(y_n-y_0-(1+\epsilon_{22})(Y_n-Y_0)-\epsilon_{21}(X_n-X_0)\right)^2.
\end{aligned}
\end{equation}
where the index $n$ runs over the particles within the sampling radius and $n=0$ for the reference particle. We then compute $\boldsymbol{\epsilon}$ for the reference particle at time $t$ that minimizes $D^2(t)$. With this method, we obtain the local strain tensors for all particles in the granular pack.    

\begin{figure*}
 \centering
 \noindent\includegraphics[width=\textwidth]{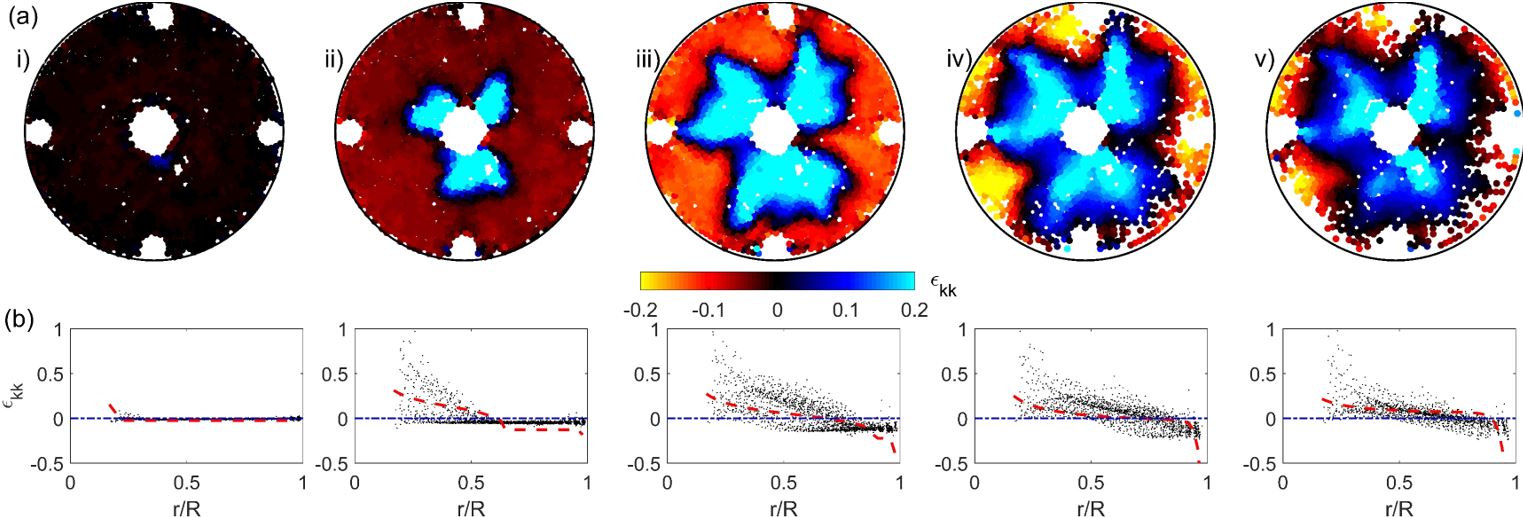}
 \caption{For the fracturing experiment with $Q=100$~mL/min, $\eta=300$~kcSt and $C=1.2\%$, we analyze the sequence of snapshots shown in Fig.~4: i), ii), iii), iv), v) corresponding to t=0.2s, 1.2s, 2.2s, 3.2s, 4.2s, respectively. The sequence of snapshots shows the time evolution of (a) experimental volumetric strain field, and (b) volumetric strain of the particles (black dots) compared with the continuum model prediction (dashed line).}
 \label{Fig6_ExpModEpsilon}
\end{figure*}

We present a sequence of snapshots of the volumetric strain field in Figure ~\ref{Fig6_ExpModEpsilon}. The granular pack dilates (positive $\epsilon_{kk}$) in the water-invaded region, and compacts (negative $\epsilon_{kk}$) in the oil-saturated region [Fig.~\ref{Fig6_ExpModEpsilon}(a)]. Such injection-induced dilation has also been reported for cohesionless granular packs and cohesive poroelastic cylinders \citep{macminn2015fluid,auton2017arteries,auton2018arteries}. Figure \ref{Fig6_ExpModEpsilon}(b) shows that the model captures the granular dilation and compaction, but cannot account for the plastic dilation near fractures brought by bond breakage and particle rearrangements.

%% Revisit water saturation curve: After breakthrough, the radial profile of water saturation are nonmonotonic, exhibiting an increase of $S_w$ with $r$ and then a decrease. The position of the local maximum of the saturation profile moves towards the center of the cell as time evolves, and the value of the maximum saturation increases with time. This unusual behavior of water saturation contrasts that of fluid-fluid displacement in a rigid porous medium \citep{buckley1942mechanism, blunt2017multiphase} and highlights the strong coupling between fluid flow and medium deformation in our system.
The injection-induced deformation also feeds back to the fluid flow, as evidenced by the observed nonmonotonic water saturation curves [Fig.~\ref{Fig4_ModPressS}(c)]. The granular pack dilation near the injection port increases the local porosity, and results in a smaller value of $S_w$. As encoded in Eqn.~\eqref{eqn_govSP_waterPress}, the coupling between fluid flow and medium deformation becomes strong when the deformation term is comparable to the flow term, $S_wb\frac{\partial\epsilon_{kk}}{\partial t}\sim\nabla\cdot \mathbf{q}_w$.

\subsection{Effective Stress}
%% Discussion on effective stress field goes as follows
The photoelastic response offers a unique opportunity to gain additional understanding of the coupled pore-scale flow and particle mechanics during fluid-induced fracturing of the cohesive granular pack. To interpret the photoelastic response, we rely on the results of calibration experiments \citep{weiLi2021}, which have shown that, for the range of interparticle forces expected in our fracturing experiments, the relation between light intensity and force is monotonically increasing and approximately linear. From two-dimensional photoelasticity theory \citep{frocht-photoelas}, the stress-optic law states that in this ``first-order'' region, the photoelastic response is approximated to be linearly proportional to the principal effective stress difference with a constant coefficient: $I=F(\sigma_1'-\sigma_2')$, where $\sigma_1'$ and $\sigma_2'$ are the maximum and minimum principal effective stresses, respectively. 

\begin{figure*}
 \centering
 \noindent\includegraphics[width=\textwidth]{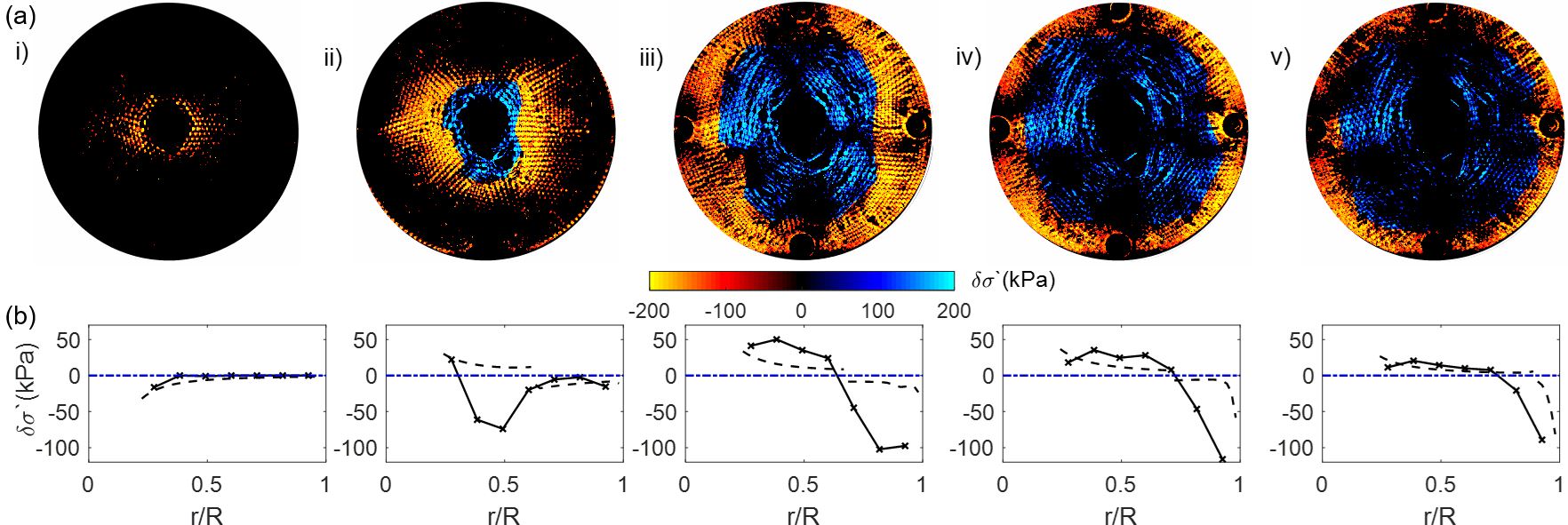}
 \caption[For the fracturing experiment with $Q=100$~mL/min, $\eta=300$~kcSt and $C=1.2\%$, we analyze the sequence of snapshots shown in Fig.~4: i), ii), iii), iv), v) corresponding to t=0.2s, 1.2s, 2.2s, 3.2s, 4.2s, respectively. The sequence of snapshots shows the time evolution of (a) experimental effective stress field, and (b) the radial distribution of the averaged effective stress (solid line) compared with the continuum model prediction (dashed line).]{For the fracturing experiment with $Q=100$~mL/min, $\eta=300$~kcSt and $C=1.2\%$, we analyze the sequence of snapshots shown in Fig.~4: i), ii), iii), iv), v) corresponding to t=0.2s, 1.2s, 2.2s, 3.2s, 4.2s, respectively. The sequence of snapshots shows the time evolution of (a) experimental effective stress field, and (b) the radial distribution of the averaged effective stress (solid line) compared with the continuum model prediction (dashed line). To differentiate the direction of force chains, we set a sign convention manually to the principal effective stress difference $\delta\sigma'$, which should otherwise always be positive, as follows: $\delta\sigma'$ is positive for tensile force chains in circumferential/hoop direction, and negative for compressive force chains in radial direction.}
 \label{Fig8_ExpMod_effStr}
\end{figure*}

To quantify the photoelastic response into the principal effective stress difference, we conduct a separate calibration experiment to obtain the coefficient, $F=0.29$ from the blue channel light intensity (see Appendix). To differentiate the direction of force chains, we set an ad~hoc sign convention for the principal effective stress difference $\delta\sigma'$, which should otherwise always be positive, as follows: $\delta\sigma'$ is positive for tensile force chains in circumferential/hoop direction, and negative for compressive force chains in radial direction. After converting $I$ into $\delta\sigma'$, and assigning its sign from the force chain direction, we present a sequence of snapshots of the effective stress field [Figure \ref{Fig8_ExpMod_effStr}(a)]. Behind the water invasion front, a \emph{hoop effective stress region}, where we observe tensile force chains in the circumferential direction, emerges and evolves as the invasion front propagates. Ahead of the invasion front, we observe radial compaction of the granular pack. 

In the model, we found that $\sigma_t'>\sigma_r'$ always holds, where $\sigma_t'$ and $\sigma_r'$ are the hoop and radial components of the effective stress, respectively. As the force chain direction aligns with the effective stress direction with larger absolute magnitude, we calculate $\delta\sigma'$ numerically with the aforementioned sign convention as follows:
\begin{equation}
   \delta\sigma'=\left\{\begin{array}{l}
   \sigma_t'-\sigma_r'>0 \text{, if } |\sigma_t'|>|\sigma_r'|,\\
   \sigma_r'-\sigma_t'<0 \text{, if } |\sigma_r'|>|\sigma_t'|.
   \end{array}\right.
 \label{eqn_effStrMod}
\end{equation}
We compare the experimental and numerical radial distribution of $\delta\sigma'$ in Fig.~\ref{Fig8_ExpMod_effStr}(b). The model captures the hoop effective stress region and radial compaction delineated by the invasion front. As mentioned in our previous discussion on the displacement field, the model cannot capture the plasticity-induced compaction front, resulting in an underestimation of compressive effective stress between times (iii) and (iv).  

\subsection{Phase Diagram of Fluid Invasion Patterns in Cohesive Granular Media}
\begin{figure*}
 \centering
 \noindent\includegraphics[height=4cm]{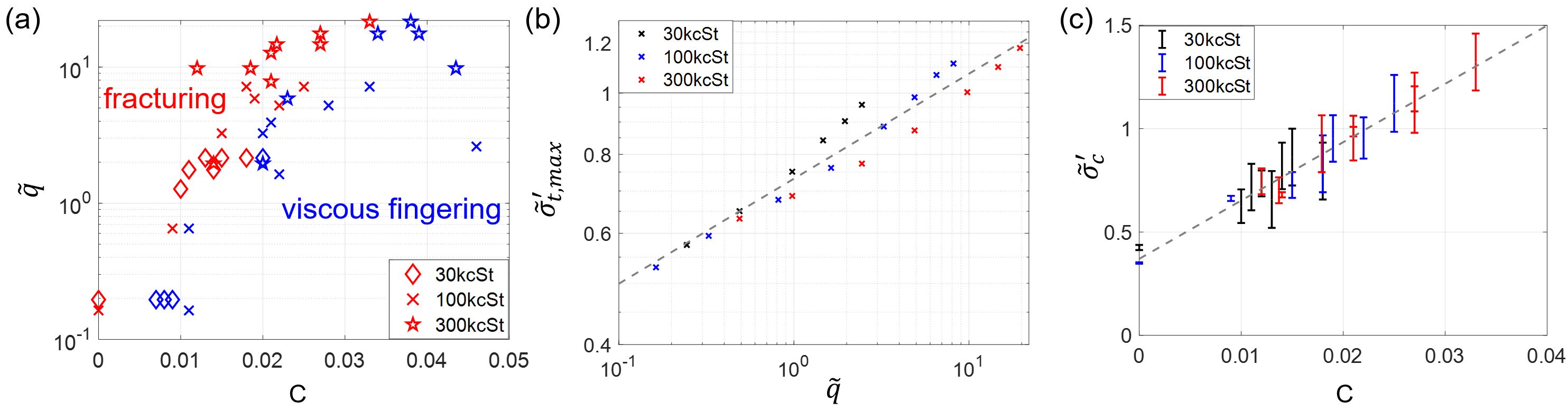}
 \caption[Phase diagrams of fluid-fluid displacement patterns in the experiments.]{Phase diagrams of fluid-fluid displacement patterns in the experiments. Diagram (a) shows the invasion patterns for all experiments, ranging in oil viscosity $\eta$ from 30~kcSt, 100~kcSt, to 300~kcSt, water injection rate $Q$ from $5$~mL/min to $220$~mL/min, and polymer content $C$ from 0 to $4.6\%$. Diagram (b) shows the modeling prediction of the dimensionless maximum effective hoop stress at the injection port, $\tilde{\sigma}'_{t,\text{max}}=\frac{\sigma'_{t,\text{max}}}{K}$, as a function of the dimensionless flow rate, $\tilde{q}=\frac{\eta Q}{2\pi k_0 MR}$. The dashed line shows the fitted power law, $\tilde{\sigma}'_{t,\text{max}}\approx0.73\tilde{q}^{0.17}$. Diagram (c) shows the dimensionless tensile strength of the granular pack against fracturing, $\tilde{\sigma}'_c=\frac{P^{\text{frac}}_{\text{inj}}}{K}$, increases with polymer content $C$. The dashed line shows the fitted linear relationship, $\tilde{\sigma}'_c=28.23C+0.37$.}
 \label{Fig9_PhaseDiag}
\end{figure*}

We observe two invasion patterns when varying the experimental parameters $\eta,Q,$ and $C$: (I) pore invasion in the form of immiscible viscous fingering, and (II) fracturing with leak-off of the injection fluid. In a granular medium, fractures open when forces exerted by the fluids exceed the mechanical forces that resist particle rearrangements. Here the competing forces are the viscous force that drives fractures, and intergranular cohesion and friction that resist fractures. For a fixed domain geometry and granular medium (particle size and packing fraction), the viscous force is expected to increase with the product of the fluid viscosity $\eta$ and the injection rate $Q$. We use a dimensionless flow rate $\tilde{q}=\frac{\eta Q}{2\pi k_0MR}$ to characterize the viscous force. The resisting force is expected to have a friction-dependent component that will be constant across our experiments, and a cohesion-dependent component that will increase with the polymer content $C$. We use a dimensionless tensile strength $\tilde{\sigma}'_c=\frac{\sigma'_c}{K}$ to characterize the resisting force. Thus, we plot an empirical phase diagram of all our experiments, indicating whether they are either ``fracturing'' or ``viscous fingering'' (not fracturing) on the axes $\tilde{q}$ vs. $C$ [Fig.~\ref{Fig9_PhaseDiag}(a)]. This empirical plot shows a transition from viscous fingering at low $\eta Q$ and high $C$ to fracturing at high $\eta Q$ and low $C$. 

In the fracturing experiment (Section 3.2), the photoelastic response reveals that fractures initiate as tensile cracks near the injection port, where intergranular bonds break under tensile stress in the circumferential direction. Shear failure also occurs during fracture propagation, as evidenced by the classic slip line fracture pattern. To rationalize the crossover from viscous fingering to fracturing regimes quantitatively, we focus on the fracture initiation and assume the tensile failure mode. We adopt a fracturing criterion for cohesive granular media: the maximum hoop effective stress ($\sigma'_{t,\text{max}}$) should exceed the tensile strength between particles ($\sigma'_c$) to break interparticle bonds and generate fractures. To theoretically predict $\sigma'_{t,\text{max}}$, we run the model with different flow conditions, and obtain $\sigma'_{t,\text{max}}$ at the injection port. We then obtain the dimensionless maximum hoop effective stress by $\tilde{\sigma}'_{t,\text{max}}=\frac{\sigma'_{t,\text{max}}}{K}$. Figure~\ref{Fig9_PhaseDiag}(b) shows that $\tilde{\sigma}'_{t,\text{max}}$ increases with $\tilde{q}$ approximately as a power law, $\tilde{\sigma}'_{t,\text{max}}\approx0.73\tilde{q}^{0.17}$. 

\begin{figure}
 \centering
 \noindent\includegraphics[height=4cm]{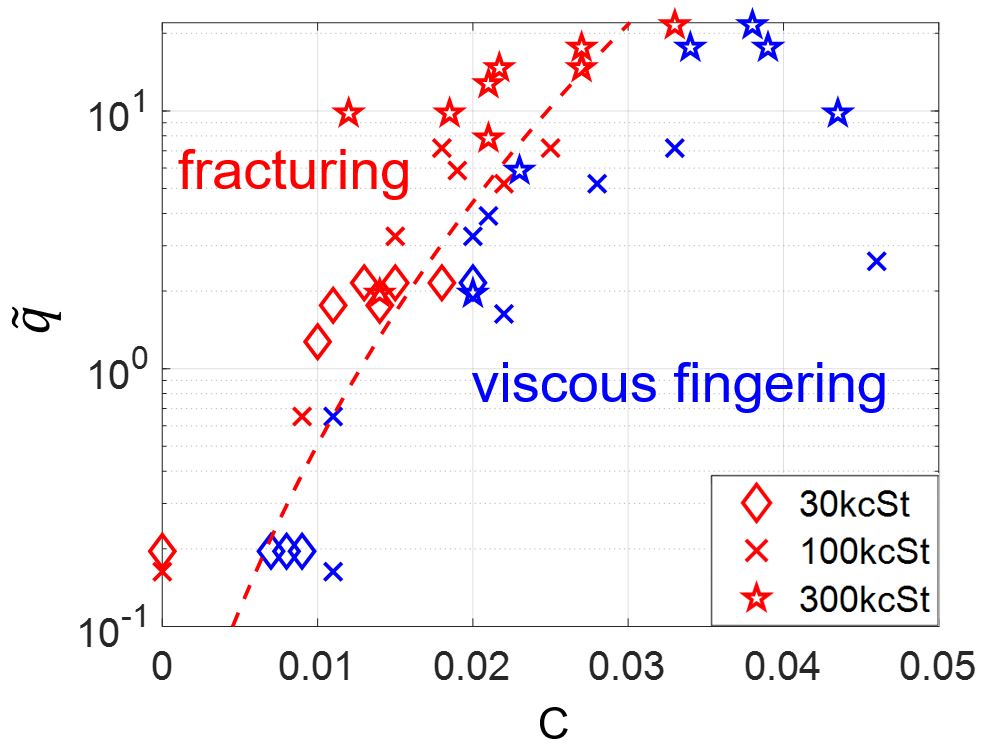}
 \caption[The experimental phase diagram of fluid-fluid displacement patterns with the theoretical prediction on the crossover from viscous fingering to fracturing regime.]{Phase diagram of fluid-fluid displacement patterns in the experiments. The dashed line is the theoretical prediction on the crossover from viscous fingering to fracturing regime, \mbox{$\tilde{q}=(38.67C+0.51)^{6.0}$.}}
 \label{Fig10_PhaseDiagThry}
\end{figure}

To construct the relationship between $\tilde{\sigma}'_c$ and $C$, we record the injection pressure at the onset of fracturing when interparticle bonds break as $P^{\text{frac}}_{\text{inj}}$. We obtain the dimensionless tensile strength $\tilde{\sigma}'_c=\frac{P^{\text{frac}}_{\text{inj}}}{K}$, and find a linear relationship, $\tilde{\sigma}'_c=28.23C+0.37$ [Figure~\ref{Fig9_PhaseDiag}(c)]. It does not pass through the origin because of the frictional resistance against fracturing for a cohesionless granular pack. After entering these relationships into the fracturing criterion, we obtain a condition involving $\tilde{q}$ and $C$ for the transition into the fracturing regime, $\tilde{q}\geq (38.67C+0.51)^{6.0}$. The theoretical prediction on the crossover from viscous fingering to fracturing regime agrees well with the experimental results [Fig.~\ref{Fig10_PhaseDiagThry}]. 

\section{Conclusions}

In summary, we have studied the morphology and rheology of injection-induced fracturing in cohesive wet granular packs via a recently developed experimental technique, photoporomechanics, which extends photoelasticity to granular systems with a fluid-filled connected pore space \citep{weiLi2021}. Experiments of water injection into cohesive photoelastic granular packs with different tensile strength, injection flow rate, and defending fluid viscosity have led us to uncover two invasion regimes: viscous fingering, and fracturing with leak-off of the injection fluid. Contrary to the observed \emph{effective stress shadow} for cohesionless granular packs \citep{yue22frac}, here we discover a \emph{hoop effective stress region} behind the water invasion front. We developed a two-phase poroelastic continuum model that captures the transient pressure response arising from the granular pack compressibility. Behind the water invasion front, the granular pack is dilated with tensile force chains in the circumferential direction. Ahead of the water invasion front, the granular pack is compacted with compressive force chains in the radial direction. Finally, we rationalize the crossover from viscous fingering to fracturing across our suite of experiments by comparing the competing forces behind the process: viscous force from fluid injection that drives fractures, and intergranular cohesion and friction that resist fractures.

The developed two-phase continuum model assumes linear elasticity, which is insufficient to capture bond breakage and particle rearrangements. In spite of its limitations, our minimal-ingredients model still sheds insight and explains some of the key features observed in the experiments. %%(summarize which experimental findings can be rationalized with the simple model you developed). 
The model reveals that the transient pressure response comes from the compressibility of the granular pack. It also captures
granular pack dilation and compaction with the boundary delineated by the invasion front, which explains the observed distinct alignments of the force chains. Furthermore, the model predicts the injection-induced hoop stress at the injection port where tensile cracks emerge, which is the key to rationalizing the crossover from viscous fingering to fracturing regimes quantitatively. 

An interesting next step would be to account for these irreversible processes by means of a poroelastoplastic or poroviscoplastic model, possibly in large deformations to reflect the substantial variations in porosity during the fluid injection. Then the poroelastic constants could be taken to be porosity-dependent. One could start with extending previous work from Auton and MacMinn \citep{auton2019large} to two-phase flow. To gain more insights on the fluid-induced fracturing, the radially symmetric model could be extended to a two-dimensional model that takes fracture morphology into account. Motivated by our experiments, Guevel et al \citep{guevel2023darcy} develop a Darcy–Cahn–Hilliard model coupled with damage to describe multiphase-flow and fluid-driven fracturing in porous media. The model adopts a double phase-field approach, regularizing both cracks and fluid-fluid interfaces. The damage model allows for control over both nucleation and crack growth, and successfully recovers the flow regime transition from fingering to fracturing with leak-off observed in our experiments. Lastly, by adding capillarity in the fluid flow equations, the model would be able to explore a wider range in $\eta$ and $Q$, and possibly explains more invasion regimes, such as capillary fingering and fracturing.     

Our study paves the way for understanding the mechanical and fracture properties of cohesive
porous materials that are of interest for applications in various fields of research and industry, such as rock mechanics \citep{dvorkin1991effect,jaeger2009fundamentals,holtzman2012micromechanical}, the fracture of concrete and biomaterials \citep{buyukozturk1998crack,topin2008wheat}, and geoengineering \citep{turcotte2014super}.

\section*{Conflicts of interest}
There are no conflicts to declare.

% ACKNOWLEDGEMENTS
\section*{Acknowledgements}
We  thank  Chris MacMinn (University of Oxford), John Dolbow (Duke University), Alex Guevel (Duke University) and Ken Kamrin (MIT) for helpful discussions. We acknowledge funding from the U.S. National Science Foundation (Grant No. CMMI-1933416).

\section*{Appendix: Calibration Experiment for Photoelastic Response}
\label{app_calExp}
The stress-optic law states that in the first order, the photoelastic response is approximated to be linearly proportional to the principal effective stress difference with a constant coefficient: $I=F\delta\sigma'$, where $\delta\sigma'$ is the principal effective stresses difference \citep{frocht-photoelas}. To obtain the coefficient $F$, we conduct a calibration experiment in the same Hele-Shaw cell where we conduct the fracturing experiments. 

\begin{figure*}
 \centering
 \noindent\includegraphics[width=\textwidth]{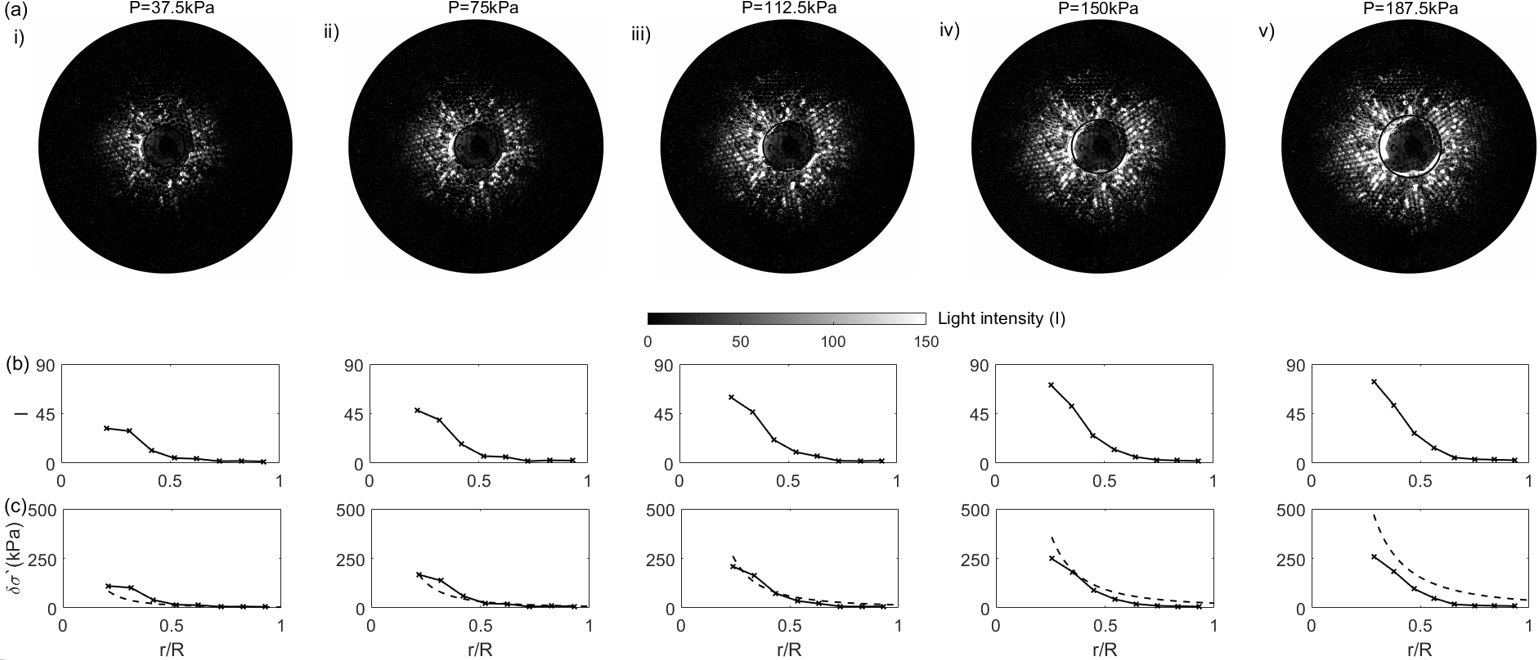}
 \caption[For the calibration experiment with increasing water injection pressure, a sequence of snapshots shows the time evolution of (a) experimental light intensity field from the blue channel, (b) the radial distribution of the averaged light intensity $I(r,t)$, and (c) the radial distribution of the averaged effective stress difference (solid line) compared with the continuum model prediction (dashed line).]{For the calibration experiment with increasing water injection pressure, a sequence of snapshots shows the time evolution of (a) experimental light intensity field from the blue channel, (b) the radial distribution of the averaged light intensity $I(r,t)$, and (c) the radial distribution of the averaged effective stress difference (solid line) compared with the continuum model prediction (dashed line). The conversion factor between light intensity and effective stress difference is calibrated to be $F=\frac{I}{\delta\sigma'}=0.29$.}
 \label{Fig7_Calib}
\end{figure*}

We prepare a monolayer of photoelastic particles at a polymer content $C=3\%$. The particle diameter and initial packing density are the same as in the fracturing experiments. We saturate the granular pack with silicone oil of viscosity 5~cSt, which lubricates the particle-particle and particle-wall contacts. After saturation, we slowly inject water at $Q=2$~mL/min into a sealed, elastic membrane that is connected to the injection port, and we monitor the injection pressure during injection. As injection proceeds, $P_{\text{inj}}$ increases and drives the outward compaction of the granular pack quasi-statically. The membrane expands in size without any water leakage. We present a sequence of snapshots of the blue-channel light intensity field from darkfield images [Fig.~\ref{Fig7_Calib}(a)]. The injection takes place under drained conditions, where the pressure in the defending fluid has time to fully dissipate, and the solid skeleton takes all the load from the water pressure at the inner boundary. The process is the same as a classical linear elastostatic example: a cylindrical vessel subject to an internal pressure and fixed outer boundary \citep{anand2020continuum}. For any specific $P_{\text{inj}}$ and size of the inner boundary, we obtain the theoretical prediction on $\delta\sigma'$, which helps us to calibrate the conversion factor $F$ between experimental light intensity [Fig.~\ref{Fig7_Calib}(b)] and effective stress difference [Fig.~\ref{Fig7_Calib}(c)]. The calibration shows that $F=0.29$ under our experimental conditions.     

%%%END OF MAIN TEXT%%%


\balance


%%%REFERENCES%%%
\bibliography{photoporofrac_cohesive} %You need to replace "rsc" on this line with the name of your .bib file
\bibliographystyle{rsc} %the RSC's .bst file

\end{document}